\documentclass[a4paper,11pt,aps,prl,superscript address]{revtex4-1}

\usepackage{graphicx, color}
\usepackage{subfigure}
\usepackage{arydshln}
\newcommand{\fig}[1]{Fig.\,\ref{#1}}
\newcommand{\Fig}[1]{Fig.\,\ref{#1}}
\newcommand{\subfig}[2]{Fig.\,\ref{#1}\,(#2)}
\newcommand{\Eq}[1]{Eq.\,(\ref{#1})}

\newcommand{\unit}[2]{$#1\,{\rm #2}$}	 	
\newcommand{\eVc}[2]{$#1\,{\rm {#2}eV}/c$}
\newcommand{\GeV}[1]{\eVc{#1}{G}}
\newcommand{\neutrons}{neutrons/cm^2}		

\newcommand{\calL}{{\cal L}}			
\newcommand{\calP}{{\cal P}}			
\newcommand{\calR}{{\cal R}}			
\newcommand{\thetaC}{\theta_{\rm C}}	
\newcommand{\Npe}{N_{\rm pe}}		

\begin{document}

\title{Particle identification performance of the prototype Aerogel RICH counter for the Belle II experiment}

\author{S.~Iwata}
\affiliation{Department of Physics, Tokyo Metropolitan University, Hachioji, Japan}

\author{I.~Adachi}
\affiliation{Institute of Particle and Nuclear Studies (IPNS), High Energy Accelerator Research Organization (KEK), Tsukuba, Japan}

\author{K.~Hara}
\affiliation{Institute of Particle and Nuclear Studies (IPNS), High Energy Accelerator Research Organization (KEK), Tsukuba, Japan}

\author{T.~Iijima}
\affiliation{Department of Physics, Nagoya University, Nagoya, Japan}

\author{H.~Ikeda}
\affiliation{Japan Aerospace Exploration Agency (JAXA), Sagamihara, Japan}

\author{H.~Kakuno}
\affiliation{Department of Physics, Tokyo Metropolitan University, Hachioji, Japan}

\author{H.~Kawai}
\affiliation{Department of Physics, Chiba University, Chiba, Japan}

\author{T.~Kawasaki}
\affiliation{Department of Physics, Kitasato University, Sagamihara, Japan}

\author{S.~Korpar}
\affiliation{Faculty of Chemistry and Chemical Engineering, University of Maribor, Maribor, Slovenia}
\affiliation{Jo\v{z}ef Stefan Institute, Ljubljana, Slovenia}

\author{P.~Kri\v{z}an}
\affiliation{Jo\v{z}ef Stefan Institute, Ljubljana, Slovenia}
\affiliation{Faculty of Mathematics and Physics, University of Ljubljana, Ljubljana, Slovenia}

\author{T.~Kumita}
\affiliation{Department of Physics, Tokyo Metropolitan University, Hachioji, Japan}

\author{S.~Nishida}
\affiliation{Institute of Particle and Nuclear Studies (IPNS), High Energy Accelerator Research Organization (KEK), Tsukuba, Japan}

\author{S.~Ogawa}
\affiliation{Department of Physics, Toho University, Funabashi, Japan}

\author{R.~Pestotnik}
\affiliation{Jo\v{z}ef Stefan Institute, Ljubljana, Slovenia}

\author{L.~\v{S}antelj}
\affiliation{Institute of Particle and Nuclear Studies (IPNS), High Energy Accelerator Research Organization (KEK), Tsukuba, Japan}

\author{A.~Seljak}
\affiliation{Department of Physics and Astronomy, University of Hawaii, Honolulu, Hawai, USA}

\author{T.~Sumiyoshi}
\affiliation{Department of Physics, Tokyo Metropolitan University, Hachioji, Japan}

\author{M.~Tabata}
\affiliation{Department of Physics, Chiba University, Chiba, Japan}

\author{E.~Tahirovic}
\affiliation{Jo\v{z}ef Stefan Institute, Ljubljana, Slovenia}

\author{Y.~Yusa}
\affiliation{Department of Physics, Niigata University, Niigata, Japan}

\begin{abstract}
We have developed a new type of particle identification device, called an Aerogel Ring Imaging Cherenkov (ARICH) counter, for the Belle II experiment.
It uses silica aerogel tiles as Cherenkov radiators.
For detection of Cherenkov photons, Hybrid Avalanche Photo-Detectors (HAPDs) are used.
The designed HAPD has a high sensitivity to single photons under a strong magnetic field.
We have confirmed that the HAPD provides high efficiency for single-photon detection even after exposure to neutron and $\gamma$-ray radiation that exceeds the levels expected in the 10-year Belle II operation.
In order to confirm the basic performance of the ARICH counter system, we carried out a beam test at the DESY using a prototype of the ARICH counter with six HAPD modules.
The results are in agreement with our expectations and confirm the suitability of the ARICH counter for the Belle II experiment.
Based on the in-beam performance of the device, we expect that the identification efficiency at \GeV{3.5} is 97.4\% and 4.9\% for pions and kaons, respectively.
This paper summarizes the development of the HAPD for the ARICH and the evaluation of the performance of the prototype ARICH counter built with the final design components.
\end{abstract}

\maketitle

\section{Introduction}
The Belle experiment \cite{B1TDR} is a B-factory experiment at the KEKB accelerator, which is an asymmetric energy $e^+e^-$ collider for the $CP$ violation search in the $B$ meson system. 
The experiment was successfully completed in 2010, and the Kobayashi--Maskawa mechanism of the $CP$ violation was confirmed.
As a next generation B-factory experiment, the KEKB accelerator and the Belle detector are being upgraded to the SuperKEKB accelerator and the Belle II detector, respectively \cite{B2TDR}.
The Belle II experiment aims to explore new physics beyond the Standard Model (BSM) through high-precision measurements of $B$ meson decays.

A new particle identification (PID) device, the Aerogel Ring Imaging Cherenkov (ARICH) counter \cite{Nishida2014}, has been developed to provide a high $\pi/K$ separation in the Belle II experiment.
The ARICH is one of the key devices for studying BSM physics e.g. measurement of $B\rightarrow\rho\gamma$ ($\rho\rightarrow\pi\pi$) decay that is contaminated with $B\rightarrow K^*\gamma$ ($K^*\rightarrow K\pi$) decay background.
This decay mode is highly suppressed in the Standard Model and there is the possibility that a non-Standard Model particle contributes in higher-order loop diagrams \cite{B2LOIphys,Arhrib2001,Akeroyd2002}.
In the Belle experiment, a threshold-type aerogel Cherenkov counter (ACC) was used for $\pi/K$ separation up to \GeV{2} in the end-cap region.
A wide momentum range up to \GeV{4} is important for the ARICH counter in Belle II, because, for example, one pion from $B\rightarrow\rho\gamma$ can have a high momentum.
The aim for the Belle II experiment is to separate kaons and pions with \unit{4}{\sigma} up to \GeV{4}.

One of the most important components of the ARICH counter is a photon detector with high position resolution.
We use a 144-ch multi-anode Hybrid Avalanche Photo-Detector (HAPD) as the photon detector.
We had studied and improved the HAPD to sustain the radiation damage during the 10-year Belle II operation of expected fluence $<10^{12}$ one MeV-equivalent $\rm \neutrons$ and dose \unit{<100}{Gy} for $\gamma$-rays.
We constructed a prototype ARICH counter using the designed HAPDs and large-size aerogel tiles.
The performance of the prototype ARICH counter is verified using a beam test as the final step of the development.

\section{Proximity-focusing ARICH counter}
\subsection{Particle identification principle}
\begin{figure}[ht]
	\centering
	\includegraphics[width=10cm]{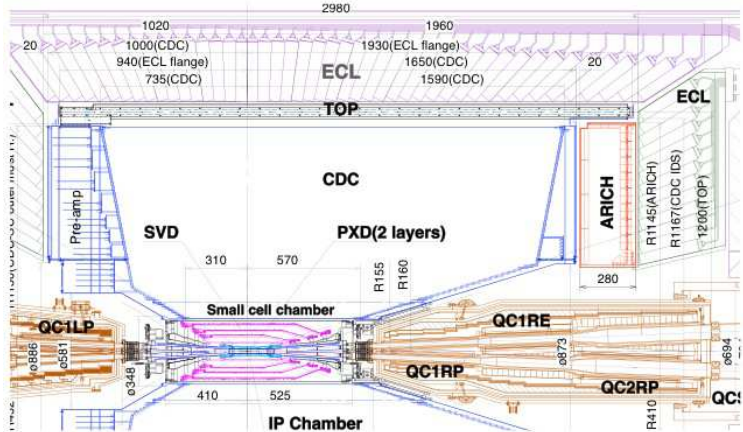}
	\caption{Horizontal cross-section of the Belle II detector.}
	\label{BelleIIdet}
\end{figure}

The ARICH counter is placed in the end-cap of the Belle II detector, as shown in \fig{BelleIIdet}.
Since the available length allowed for the ARICH counter is limited to \unit{280}{mm} along the beam line, we developed a proximity-focusing Ring Imaging Cherenkov (RICH) detector consisting of aerogel radiators, photon detectors with high efficiency and position resolution, and compact readout electronics.

\begin{figure}[ht]
	\centering
	\includegraphics[width=7cm]{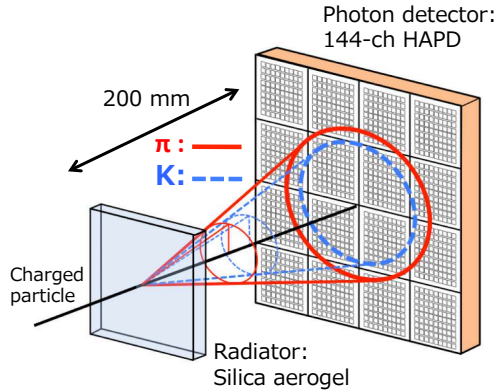}
	\caption{The principle of $\pi/K$ identification for the ARICH counter.
	The solid-line and dotted-line cones illustrate the emitted Cherenkov light for a pion and a kaon, respectively.}
	\label{prin_ARICH}
\end{figure}

\Fig{prin_ARICH} shows the particle identification principle of an ARICH.
When a charged particle is traveling through the silica aerogel radiator, Cherenkov light is produced if the velocity of the particle exceeds the speed of light in the radiator.
The Cherenkov photons are emitted at a certain Cherenkov angle with respect to the direction of the incident particle, and can be detected as a ring image by setting photon detectors as an imaging device at some distance from the radiator.
Using the Cherenkov angle $\thetaC$ calculated from the radius of the Cherenkov ring, particle identification of a charged particle can be performed by calculating its mass using the following formula:
\begin{eqnarray}
	m=\frac{p}{c}\sqrt{n^{2}\cos^{2}\thetaC-1},
	\label{Cherenkov_fomula}
\end{eqnarray}
where $p$ is the particle momentum measured by tracking the particle in a magnetic field, $c$ is the speed of light in vacuum, and $n$ is the refractive index of the aerogel. 
A pion and kaon with the same momentum emit Cherenkov light with different Cherenkov angles.
The difference of the Cherenkov angle between pion and kaon is around \unit{23}{mrad} for \GeV{p=4} and $n=1.05$.

\subsection{Aerogel radiators}
The silica aerogel radiator tiles of the ARICH counter are required to have a long transmission length and refractive indices in the range of 1.04--1.05 for optimal counter performance in the required momentum range. 
We successfully established a method for large-size (\unit{180 \times 180 \times 20}{mm^3}) hydrophobic aerogel production with high transparency~\cite{Tabata2012,Tabata2014}.

\begin{figure}[ht]
	\center
	\subfigure[Single-layer system]{
		\includegraphics[width=0.3\columnwidth]{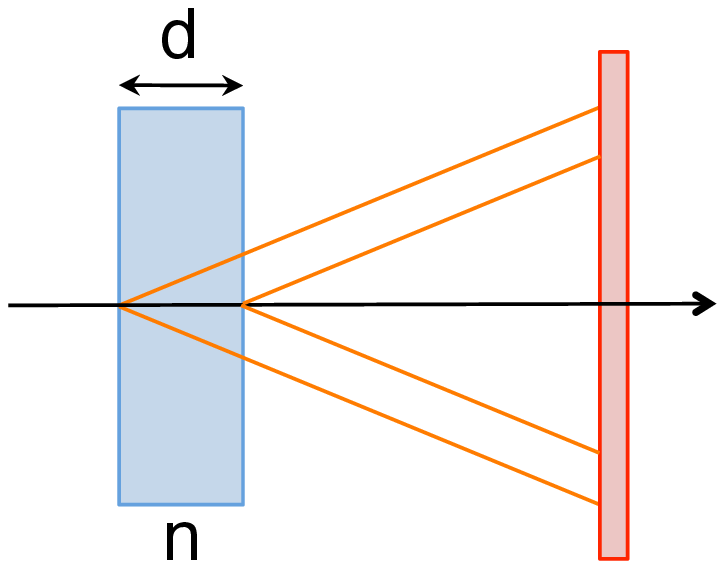}
	}
	\subfigure[Dual-layer system]{
		\includegraphics[width=0.3\columnwidth]{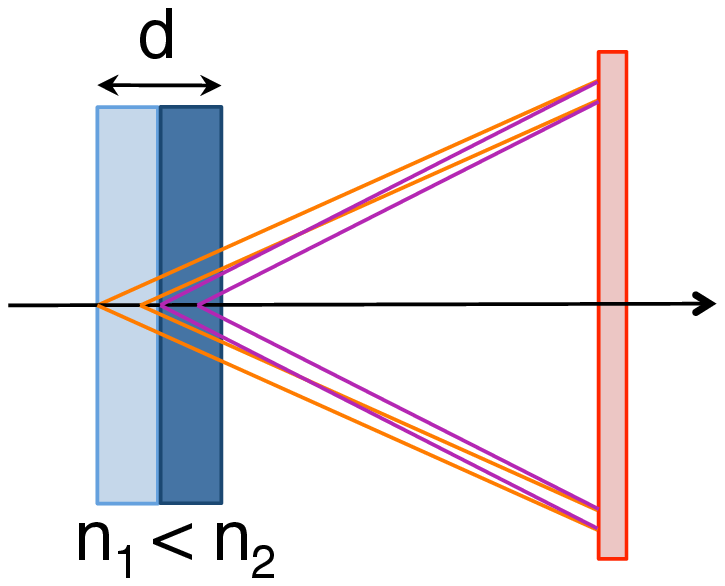}
	}
	\caption{The proximity dual-layer focusing scheme:
	(a) image of a normal Cherenkov counter with a single layer;
	(b) proximity-focusing type with dual layer, in which different refractive indices of $n_1$ and $n_2$ ($n_1 < n_2$) are used. 
	The total thickness of the aerogel(s) $d$ for both systems is the same.}
	\label{aerogel}
\end{figure}

The single-photon Cherenkov angle resolution and the number of detected Cherenkov photons per track are important parameters for high-precision measurement of particle velocity (average Cherenkov angle) with the ARICH counter.
For a normal proximity-focusing RICH counter [\subfig{aerogel}{a}] the contribution of the radiator thickness to the resolution of the average Cherenkov angle $\sigma_d$ is proportional to $d/\sqrt{\Npe}$, where $d$ is the thickness of the radiator and $\Npe$ is the number of detected photons in the ring.
If the absorption length is large compared to $d$ the number of detected photons increases linearly with the thickness and $\sigma_d$ is proportional to $\sqrt{d}$.
The thinner aerogel can improve $\sigma_d$ due to the decreased uncertainty of the emission point of a Cherenkov photon, although the detected photons will be decreased.
We verified that the optimal thickness of the aerogel for the ARICH counter should be around \unit{20}{mm}~\cite{Matsumoto2004,Iijima2005}.

In order to increase the number of detected photons without degrading the resolution, we introduced the dual-layer focusing scheme \cite{Iijima2005}.
We use two aerogel tiles with different refractive indices placed together [\subfig{aerogel}{b}].
By adjusting the refractive index of the downstream aerogel $n_2$ to be slightly higher than that of the upstream aerogel $n_1$, the Cherenkov angle in the downstream aerogel becomes slightly larger than that of the upstream aerogel.
As a result, the Cherenkov photons from the two aerogel tiles are focused on the incident windows of the photon detectors.
Here, $n_1$ and $n_2$ are 1.045 and 1.055, respectively \cite{Iijima2005}.

\subsection{Photon detector}
The Cherenkov photons are detected using an array of position-sensitive photon detectors that are located about \unit{200}{mm} downstream of the aerogel radiator.
Because the difference in radii of the Cherenkov rings between a pion and kaon at \GeV{4} is only \unit{5}{mm}, we require a pixel size of \unit{5}{mm} to separate a Cherenkov ring of them.
The photon detector has a position resolution of \unit{5/\sqrt{12}}{mm}.
The photon detector is required to have the following characteristics:
\begin{enumerate}
	\item compact in height perpendicular to the photo-cathode,
	\item pixelated anodes with pixel size of about \unit{5 \times 5}{mm^2},
	\item excellent sensitivity for single photons,
	\item the capability to operate in a high magnetic field of \unit{1.5}{T}.
\end{enumerate}
For this purpose, we have been developing a 144-ch multi-anode HAPD together with Hamamatsu Photonics K.K. since 2002 \cite{Adachi2010,Korpar2014}.
We will use 420 HAPDs for the ARICH counter.
The details of the HAPD will be described in Sect. 3.

\subsection{Readout electronics}
The readout electronics of the ARICH counter are required to have high gain and a low-noise amplifier so that it can discriminate the single photon signal from noise.
Note that the charge information is used only for discrimination between single photons and noise; most of the Cherenkov photon hits are from single photons, and only the hit information (yes or no) is important.
The readout electronics are also required to fit within the limited space; the space available for the electronics is only \unit{50}{mm} of the total available space of \unit{280}{mm}.

In order to satisfy these conditions, we developed a custom ASIC, named SA03 \cite{Kakuno2014}.
The ASIC has 36 channels with a charge-sensitive amplifier, shaper, and discriminator.
The shaper has a variable shaping time that allows the readout electronics to adapt to operations in the case of an increasing noise level due to neutron irradiation.
The details of this topic will be described in Sect. 3.

The front-end board, which is attached to the backplane of an HAPD, has four readout ASIC chips and an FPGA chip (Spartan6) for readout control and communication to the higher levels of the readout system.
The other components on the front-end board are a bias voltage connector for the attached HAPD and temperature sensor chips \cite{Seljak2}.
The size of the front-end board is designed to fit the HAPD.

\section{Hybrid Avalanche Photo-Detector}
\subsection{Specification}
\begin{figure}[ht]
	\centering
	\includegraphics[width=10cm]{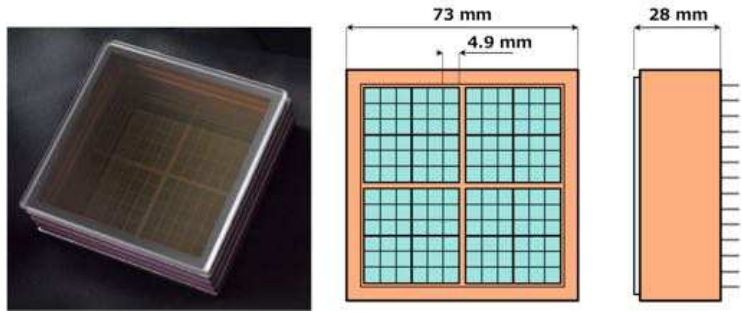}
	\caption{Picture of the exterior and the design of the 144-ch HAPD.}
	\label{HAPD}
\end{figure}

The HAPD is composed of a photo-cathode, a vacuum tube, and four avalanche photo-diode (APD) chips, where each APD is pixelated into $6 \times 6$ pads, resulting in 144 channels.
The schematics of the HAPD are shown in \fig{HAPD}.
The basic specifications of the HAPD are summarized in Table~\ref{HAPDspec}.

\begin{figure}[ht]
	\center
	\subfigure[A cross-sectional view of the HAPD]{
		\includegraphics[width=8.5cm]{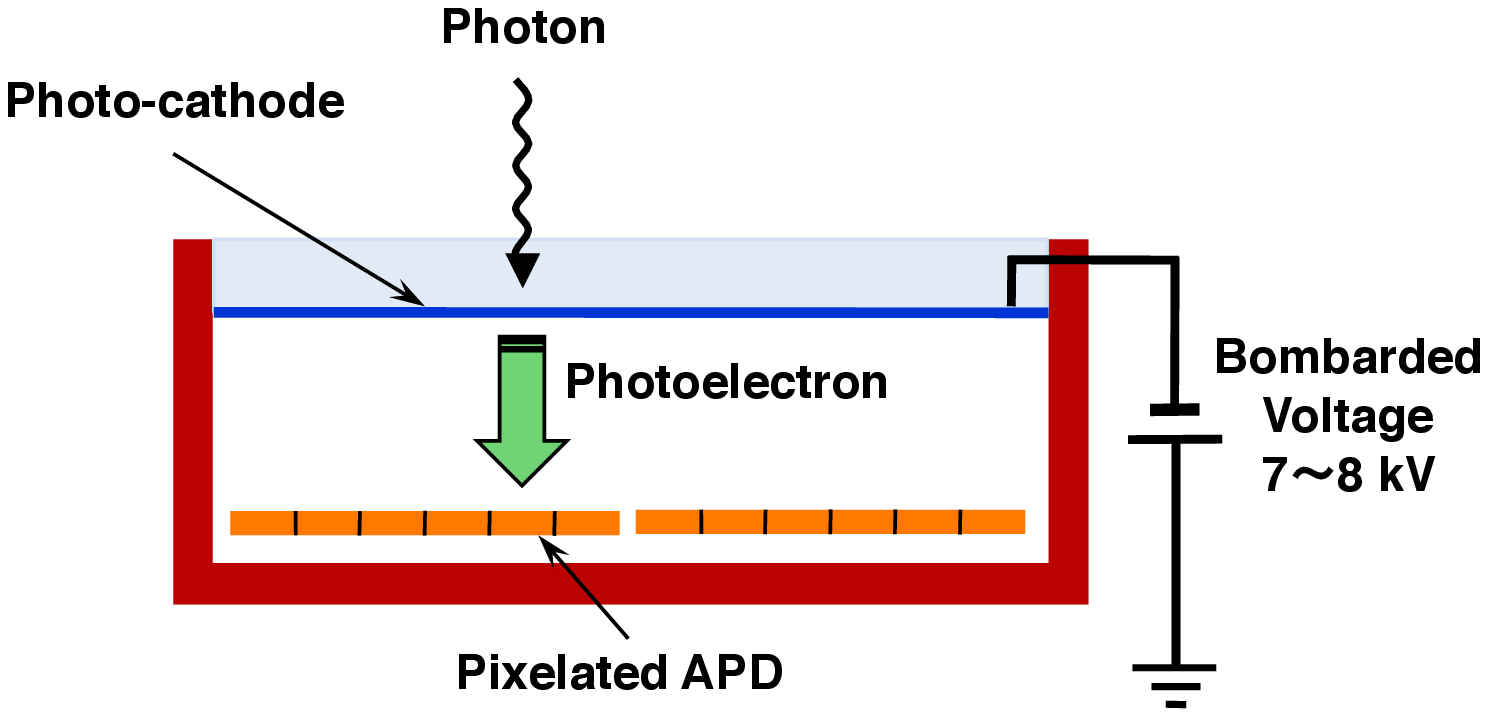}
	}
	\subfigure[A illustration of the structure of an APD]{
		\includegraphics[width=5.5cm]{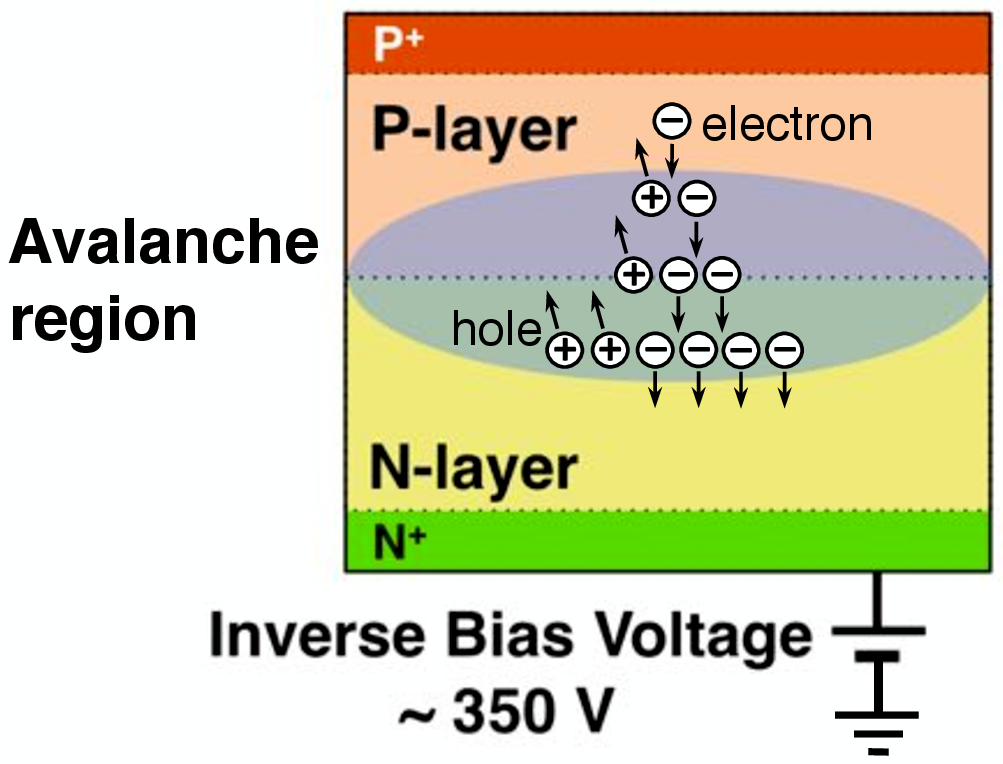}
	}
	\caption{Illustrations of the HAPD and the structure of an APD.
	There are two mechanisms of electron amplification.
	These include bombardment gain due to the electric field (a), and avalanche gain in the APD (b).}
	\label{HAPDcrosssection}
\end{figure}

\begin{table}[ht]
	\renewcommand{\arraystretch}{1.3}
	\caption{HAPD specification}
	\label{HAPDspec}
	\centering
	\begin{tabular}{l c}
		\hline \hline
		\# of pixels & $12 \times 12$ = \unit{144}{ch}\\
		Package size & \unit{73 \times 73 \times 28}{mm^3}\\
		Pixel size & \unit{4.9 \times 4.9}{mm^{2}}\\
		Effective area & 65\%\\
		Capacitance & \unit{80}{pF}\\ \hline
		Window material & Synthetic quartz \\
		Window thickness & \unit{3}{mm} \\
		Photo-cathode material & Bialkali \\
		Quantum efficiency & $\sim28\%$ (average, @\unit{400}{nm})\\ \hline
		Avalanche gain & $\sim 40$ (usually)\\
		Bombardment gain & $\sim 1700,~ @\unit{7}{kV}$ \\
		$S/N$ & $\sim 15$\\
		\hline \hline
	\end{tabular}
\end{table}

\subsection{Detection of single photon}
In an HAPD module, the photoelectrons are amplified in two steps (\fig{HAPDcrosssection}).
In the first step, the photoelectrons are accelerated using a high electric field; after passing a potential difference of 7--\unit{8}{kV} in vacuum it hits the APD, and produces about 1700 electron--hole pairs [\subfig{HAPDcrosssection}{a}].
This gain is known as bombardment gain.
In the second step, avalanche amplification occurs in the APD.
The generated electron produces around 40 electron--hole pairs in the high-filed region of the APD with an inverse bias voltage of around \unit{350}{V} [\subfig{HAPDcrosssection}{b}].
This gain is known as avalanche gain.
As a result, the total gain becomes around $7 \times 10^4$.

\begin{figure}[ht]
	\centering
	\includegraphics[width=9cm]{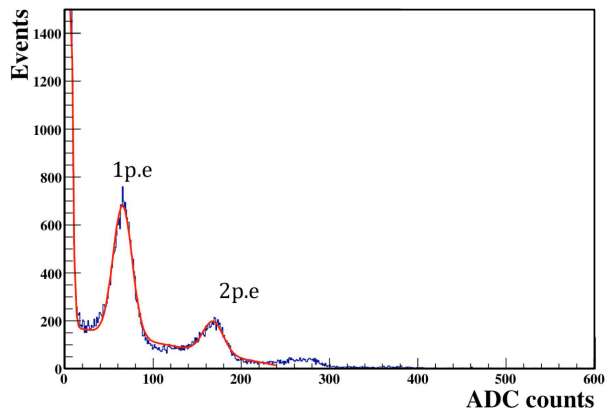}
	\caption{The pulse height distribution for low-intensity pulsed LED illumination.}
	\label{HAPD_phd}
\end{figure}

Because the bombardment gain [${\sim \cal O}(10^3)$] is larger than the avalanche gain, the statistical fluctuation in the output signal can be suppressed; this effect is further enhanced using the Fano factor in silicon.
Therefore, the HAPD has excellent performance in single photoelectron separation.
\Fig{HAPD_phd} shows the pulse height distribution with multiple photons produced using a blue LED.
We fit this distribution with a sum of three Gaussians for the three peaks, corresponding to the noise, single photoelectrons (\unit{1}{p.e.}), and two photoelectrons (\unit{2}{p.e.}), and a second-order polynomial function.
From the difference of the mean value between \unit{1}{p.e.} and \unit{2}{p.e.}, the signal gain of \unit{1}{p.e.} was calculated to be about 46,000.
The noise equivalent to the number of electrons was calculated as about 2000 from the width (\unit{1}{\sigma}) of the leftmost Gaussian.
We have therefore obtained a signal-to-noise ratio $S/N$ of about 23, sufficient for reliable single-photon detection in the ARICH counter.

\subsection{Radiation hardness}
Besides the mechanical and electrical constraints, the HAPD is also required to have sufficient radiation tolerance for the 10-year Belle II operation. 
In the front of the end-cap region of Belle II, a one MeV-equivalent neutron fluence of \unit{1 \times 10^{12}}{cm^{-2}} and a $\gamma$-ray dose of up to approximately \unit{100}{Gy} are expected in total for the 10 year of operation. 

In general, neutrons induce lattice defects in the bulk region of an APD.
This results in an increase of the leakage current through this region.
The shot noise from the HAPD also increases and causes a degradation of $S/N$ in the single-photon detection. 
We had performed several irradiation tests to investigate the dominant source of shot noise.
As a result, we confirmed that the leakage current from the P-layer is larger than that from the N-layer because the electrons from the P-layer get amplified in the avalanche region, while the amplification for holes from the N-layer is negligible [\subfig{HAPDcrosssection}{b}].
The leakage current can therefore be more efficiently reduced with a thinner P-layer, and this solution to suppress the noise increase due to bulk damage was implemented in the final version of the APD. 

The noise can also be suppressed by a shorter shaping time in the front-end electronics.
The main components of the noise for each channel of the HAPD are the shot noise and amplification noise.
We assume that the shot noise and amplification noise are proportional to $\sqrt{I_{\rm leak}G\tau}$ and $1/\sqrt{\tau}$, respectively \cite{Iwata2012}. 
Here, $I_{\rm leak}$ is the leakage current, $G$ is the avalanche gain, and $\tau$ is the shaping time.
Because the shot noise is proportional to $\sqrt{\tau}$, it can be suppressed by using a shorter shaping time.
However, the total noise increases for very short shaping times, as the amplification noise is inversely proportional to $\sqrt{\tau}$.
We have calculated the optimal shaping time, which is around \unit{100}{ns} with $G = 40$.
For this purpose, we developed a readout ASIC, where the shaping time can be varied between \unit{100}{ns} and \unit{200}{ns}.

The $\gamma$-ray radiation causes charge up around the structure on the APD surface.
In particular, we found that a protection film, which is deposited on the APD in order to protect it from alkali materials that are evaporated during photo-cathode deposition, was easily charged up by $\gamma$-rays, and the breakdown voltage between the film and the structure around it was reduced as the $\gamma$-ray dose was increased.
As a result, the breakdown voltage is reduced below the normal operational voltage, and the APD has to be operated at lower avalanche gain.
Therefore, the surface part of the APD had to be redesigned to prevent the charge up, while maintaining sufficient protection from alkali materials.

In 2012,  we performed a series of irradiation tests to confirm that the HAPD with an improved APD has a sufficient tolerance to both neutron and $\gamma$-ray irradiations expected at Belle II \cite{Iwata2012}. 
We first performed the neutron irradiation test at the neutron beam line in J-PARC MLF (Ibaraki, Japan).
The HAPDs having a thinner P-layer were irradiated with a fluence of up to \unit{0.86 \times 10^{12}}{\neutrons} (one MeV-equivalent).

\begin{figure}[ht]
	\centering
	\includegraphics[width=10cm]{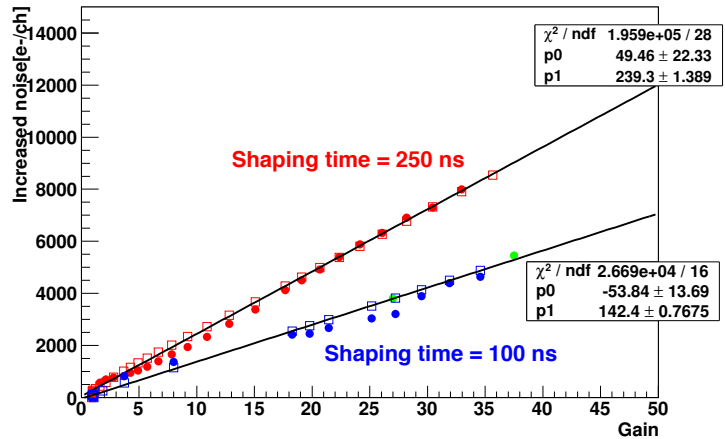}
	\caption{Results of the neutron irradiation test in 2012. 
	The measured noise is plotted with filled symbols as a function of the effective avalanche gain corresponding to a reduced bias voltage due to the increased leakage current;
	the filled symbols correspond to measurements with \unit{250}{ns} (red) and \unit{100}{ns} (blue) after irradiation with neutrons corresponding to a fluence of \unit{0.86 \times 10^{12}}{cm^{-2}}.
	The open symbols correspond to estimated noise levels assuming the shot noise for \unit{250}{ns} and \unit{100}{ns}, and the noise expectations are fitted with the solid line.
	}
	\label{neutron_test}
\end{figure}

\Fig{neutron_test} shows the result of the neutron irradiation test for the HAPD irradiated with \unit{0.86 \times 10^{12}}{\neutrons}.
This plot compares the noise levels for shaping times of \unit{100}{ns} and \unit{250}{ns}.
We confirmed that shaping time of \unit{100}{ns} suppresses the noise induced by the neutron irradiation, and the noise levels for \unit{100}{ns} is lower than in the case of \unit{250}{ns}.
The noise after neutron irradiation is expected to be around \unit{5,000}{e^-} with an avalanche gain of 40.
Since the single photoelectron signal is estimated as $1700 \times 40 = 68,000 ({\rm e^-})$ with nominal operational bombarded voltage and inverse bias voltage, we estimate the $S/N$ to be greater than 10.
Such a value is acceptable for the ARICH counter.
As a result, it can be concluded that the developed HAPD sensor, which has a thinner P-layer and is read out with shorter shaping time, can separate single photoelectrons from noise even after the expected level of neutron irradiation.

\begin{figure}[ht]
	\centering
	\includegraphics[width=10cm]{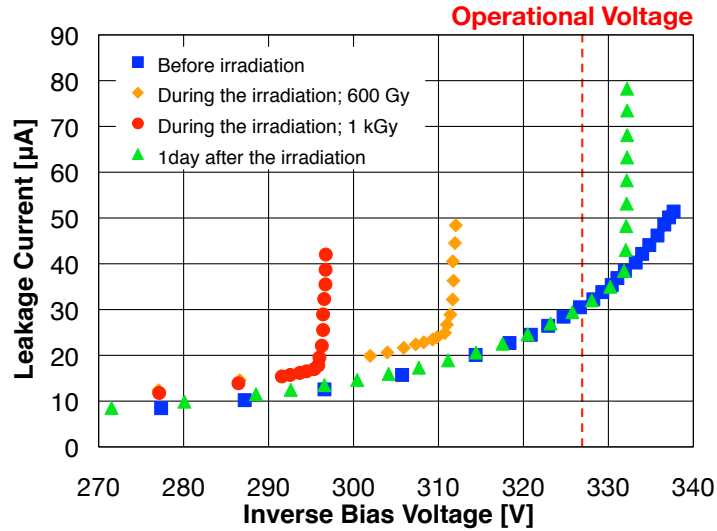}
	\caption{Comparison of leakage current before and after the irradiation:
	before the irradiation (blue square), after \unit{600}{Gy} (orange rhombus) and \unit{1,000}{Gy} (red circle), and one day after irradiation (green triangle).
	The operational voltage is indicated as a vertical dashed line corresponding to \unit{327}{V}.}
	\label{gamma_test}
\end{figure}

Around three months after the neutron irradiation, a $\gamma$-ray irradiation test for doses up to \unit{1,000}{Gy} was performed for all the neutron-irradiated HAPDs at a $^{60}$Co facility at Nagoya University in 2012.
A comparison of the leakage current before and after the $\gamma$-ray irradiation is shown in \Fig{gamma_test}.

If measured immediately after the irradiation, the breakdown voltages for the APDs with $\gamma$-ray doses of \unit{600}{Gy} and \unit{1,000}{Gy} are below the operational voltage of the APD (\unit{327}{V}, corresponding to an avalanche gain of 40).
However, after a day of annealing at room temperature, the breakdown voltage again exceeds the operational voltage.
This indicates that the effect is due to the charging up of the surface at irradiation rates exceeding by several orders of magnitude the rates expected in Belle II.
It is therefore expected that the charging up will eventually pose no problem at the expected total doses of up to \unit{100}{Gy}.
In summary, we have confirmed that the HAPDs will reliably operate up to the maximal expected neutron fluences and $\gamma$-ray doses, and even beyond.

\section{Beam test with the prototype ARICH}
In order to confirm the basic performance of the ARICH counter using the developed components including a neutron and $\gamma$-ray irradiated HAPD, we constructed a prototype detector and have carried out a test at the electron beam line T24, which provides an electron beam at DESY.

\subsection{Prototype ARICH}
The prototype ARICH counter consists of six HAPD modules, two aerogel tiles, and six front-end boards.
The specifications of the aerogels and HAPDs used are listed in Table\ref{spec_protoARICH}, and the HAPD layout is shown in \subfig{exp_setup}{a}.
The two aerogel layers were mounted in front of the HAPD array.

HAPD No. 4 was used for the radiation hardness tests, irradiated with a neutron fluence of \unit{0.86\times10^{12}}{cm^{-2}} and with a $\gamma$-ray dose of \unit{1,000}{Gy}.
HAPD signals were read out by the front-end boards attached to the backplanes of the HAPD modules.
The shaping times of the ASICs in the front-end boards were set to be \unit{100}{ns} for HAPD Nos. 2, 4, and 6, and \unit{250}{ns} for Nos. 1, 3, and 5. 

\begin{table}[ht]
	\renewcommand{\arraystretch}{1.2}
	\caption{Basic specification of the prototype ARICH; 
	$n$ is the refractive index, $\Lambda_T$ is the transmission length and $d$ is the thickness for an individual aerogel tile.}
	\label{spec_protoARICH}
	\centering
	\begin{tabular}{c c c c c c}
		\hline \hline
		Aerogels & Position & $n$ & $\Lambda_T$ & $d$ & dimensions\\
		& Upstream & 1.0467 & \unit{47}{mm} & \unit{20.3}{mm} & \unit{182 \times 182}{mm^2}\\
		& Downstream & 1.0592 & \unit{59}{mm} & \unit{20.3}{mm} & \unit{168 \times 168}{mm^2}\\
		\hline
		HAPDs & ID & QE(peak) & \multicolumn{3}{c}{Remarks}\\
		& No. 1 & 27.4\% &&&\\	
		& No. 2 & 25.2\% &&&\\	
		& No. 3 & 28.9\% &&&\\	
		& No. 4 & 31.1\% & \multicolumn{3}{c}{neutrons and $\gamma$-ray irradiated} \\	
		& No. 5 & 26.8\% &&&\\	
		& No. 6 & 22.3\% &&&\\	
		\hline \hline
	\end{tabular}
\end{table}

\subsection{Experimental setup}
\begin{figure}[ht]
	\center
	\subfigure[The HAPD layout]{
		\includegraphics[width=4cm]{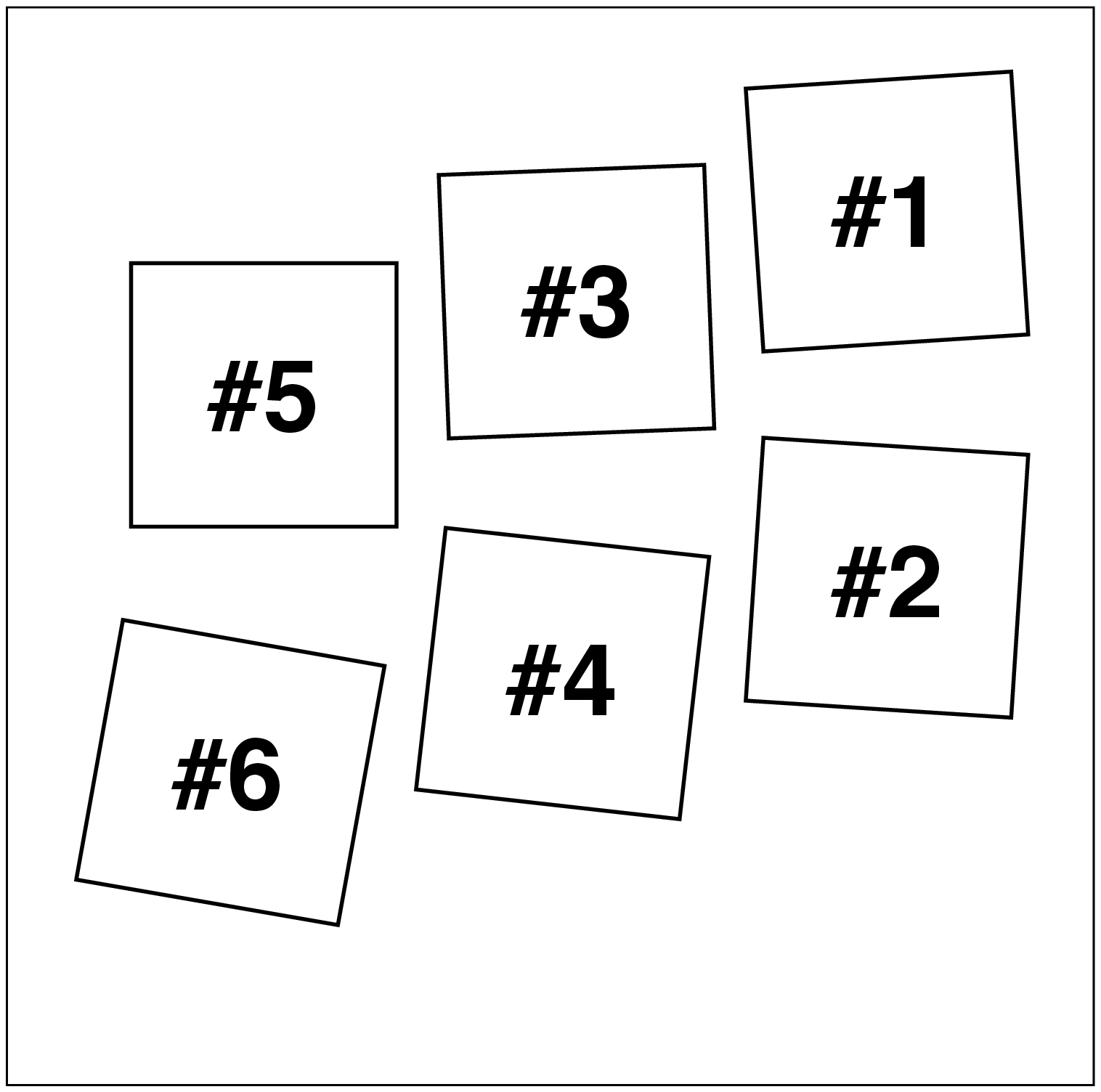}
	}
	\subfigure[The experimental setup]{
		\includegraphics[width=9cm]{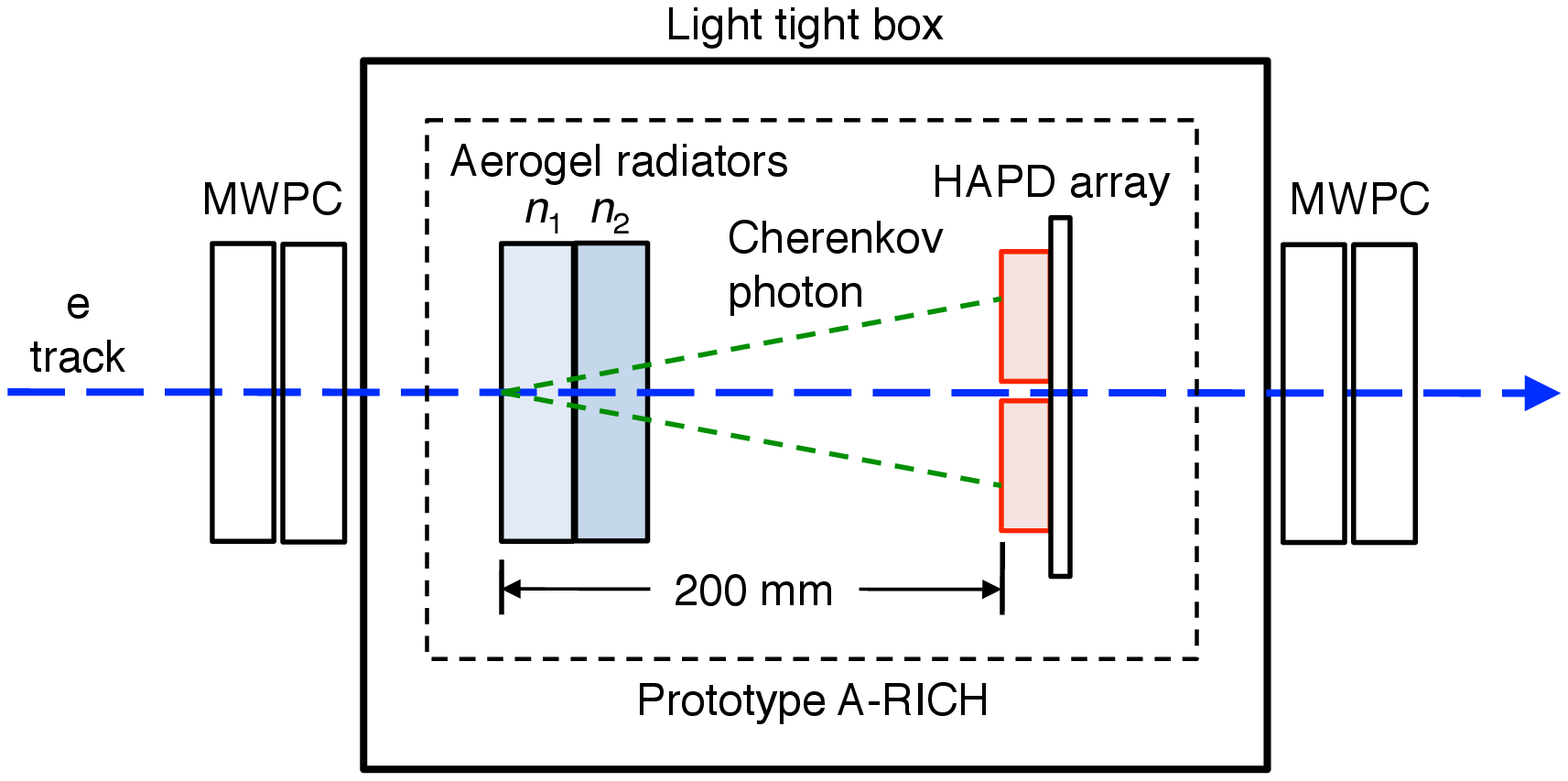}
	}
	\caption{The HAPD layout of the prototype ARICH counter and a cross-sectional view of the experimental setup of the beam test.}
	\label{exp_setup}
\end{figure}

Figure\,\ref{exp_setup}(b) shows the experimental setup of the test.
We used four multi-wire proportional chamber (MWPC) modules as the tracking device, and a pair of plastic scintillation counters for trigger generation. 
They were arranged in the front-end and rear-end of the light-tight box housing the prototype counter.
The beam direction was perpendicular to the photo-detector and aerogel planes. 
All the tests were carried out in the absence of a magnetic field.

\subsection{Photoelectron yield}
\begin{figure}[ht]
	\center
	\subfigure[The event display example]{
		\includegraphics[width=6.5cm]{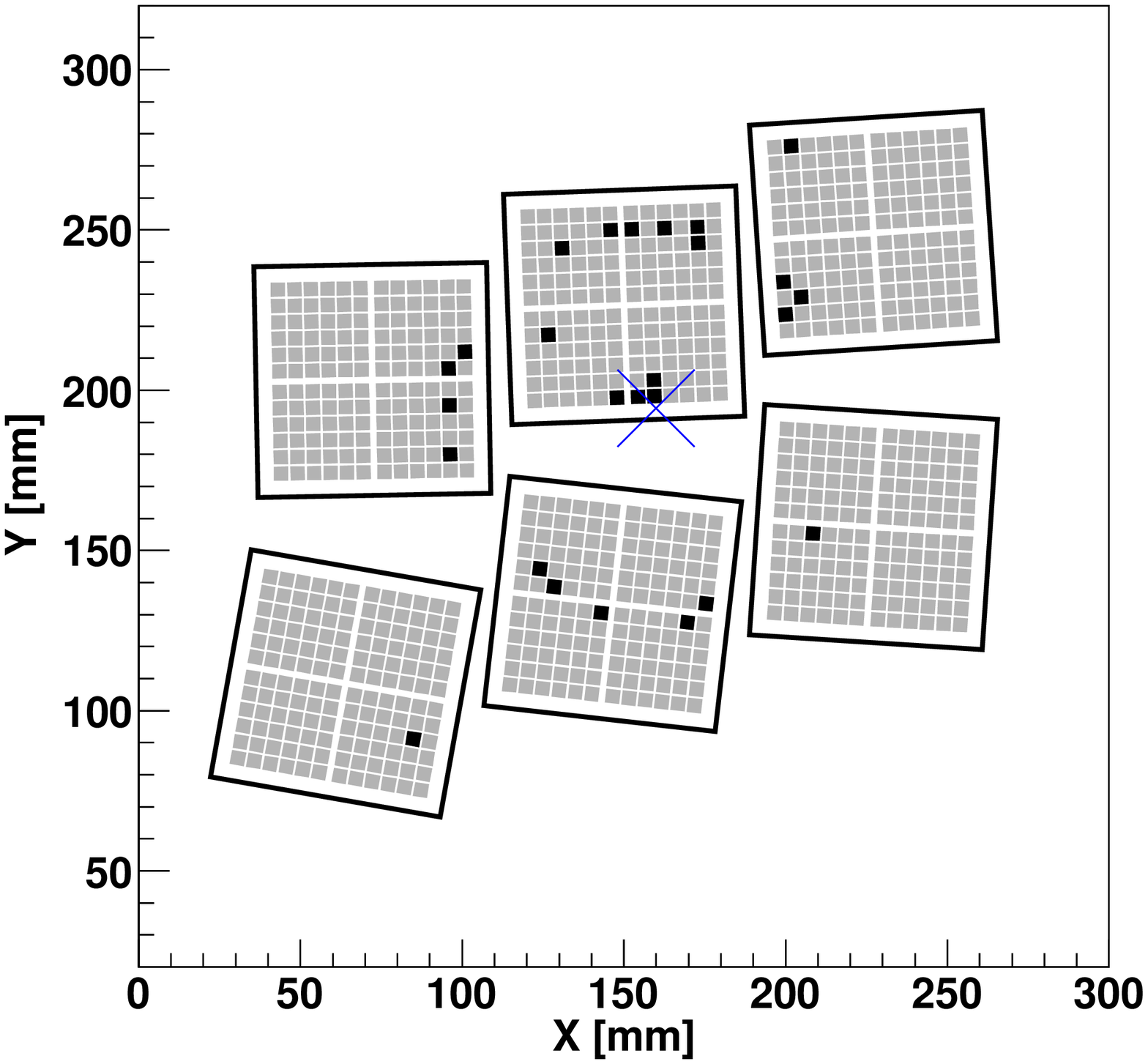}
	}
	\subfigure[The accumulated hit positions on the HAPD plane]{
		\includegraphics[width=7.3cm]{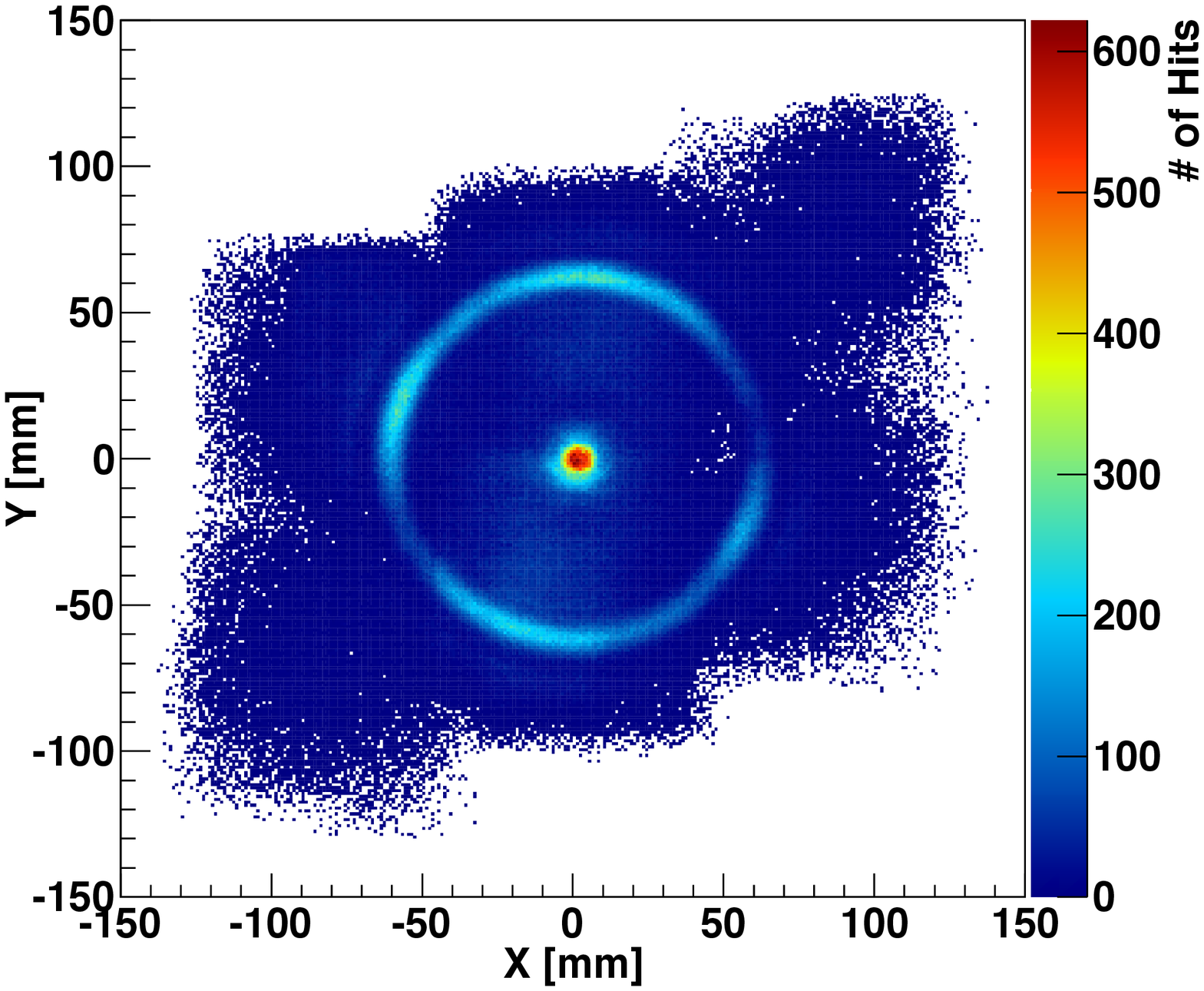}
	}
	\caption{Example of the event display and accumulated Cherenkov ring image.}
	\label{Cherenkov_ring}
\end{figure}

\Fig{Cherenkov_ring} shows the Cherenkov ring images from the prototype, a typical hit map of an event and the accumulated hit positions with respect to the track.
In \subfig{Cherenkov_ring}{a}, the cross marker corresponds to the track position; several hits seen around it correspond to Cherenkov photons which were mainly generated in the front quartz window of the HAPD.
We successfully observed very clear ring images using the prototype ARICH counter including the irradiated HAPD.

In order to evaluate the basic performance of the ARICH, we analyzed the number of detected Cherenkov photoelectrons and their angular resolution.
Multi-track events were rejected by using information from trigger counters and MWPC modules.
Those selections at the hardware level, however, cannot fully reject multi-track events.
We also applied an analytical selection to further reject such events in offline analysis.

\begin{figure}[ht]
	\centering
	\includegraphics[width=7cm]{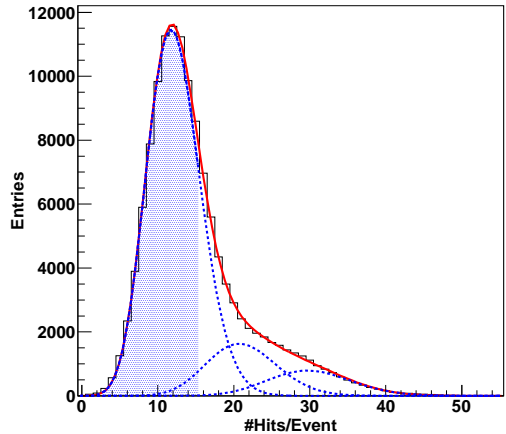}
	\caption{Distribution of the number of detected photons per event.
	The solid line (red) represents the fitted function [\Eq{Npe_fitter}], and the dotted lines (blue) show decomposed Poisson functions corresponding to single-, double-, and triple-track events.
	}
	\label{numhits}
\end{figure}

\Fig{numhits} shows the distribution of the number of detected Cherenkov photons per event.
Only photons within the ring area with Cherenkov angle in the range of \unit{\pm 45}{mrad} around the expected Cherenkov angle were counted.
Here, \unit{45}{mrad} equals \unit{3}{\sigma} of the Cherenkov photon angular distribution obtained by the fit.

In order to separate single-track events from multi-track events,  we assumed that the distribution (\fig{numhits}) was represented by the following form:
\begin{eqnarray}
	A_1 Po(N+B) + A_2 Po(2N+B) + A_3 Po(3N+B),
	\label{Npe_fitter}
\end{eqnarray}
where $Po(x)$ is the Poisson function, $A_i$ are coefficients corresponding to the number of events with $i$ tracks, $N$ is the number of detected Cherenkov photons per track, and $B$ is the average number of uncorrelated background hits per event.

the expected number of detected photoelectrons is given by the following equations \cite{B2TDR}:
\begin{eqnarray}
	\label{Npe_estimation}
	\Npe &=& N_1 + N_2,\\
	N_1 &=& 2\pi\alpha\ \sin^2\theta_{\rm C1}\epsilon_a\int\exp\left( -\frac{d_2}{\Lambda_2(\lambda)\cos\theta_{\rm C1}} \right) \nonumber \\
		&&\times \Lambda_1(\lambda)\cos\theta_{\rm C1}\left( 1-\exp\left( -\frac{d_1}{\Lambda_1(\lambda)\cos\theta_{\rm C1}} \right) \right) \epsilon_q(\lambda)\lambda^{-2}d\lambda,\nonumber \\
	N_2 &=& 2\pi\alpha\ \sin^2\theta_{\rm C2}\epsilon_a\int \Lambda_2(\lambda) \cos\theta_{\rm C1} \left( 1-\exp\left( -\frac{d_2}{\Lambda_2(\lambda)\cos\theta_{\rm C2}} \right) \right) \epsilon_q(\lambda)\lambda^{-2}d\lambda, \nonumber
\end{eqnarray}
where $N_1$ ($N_2$) is the number of photoelectrons due to Cherenkov light emitted in the upstream (downstream) aerogel tile,
$\epsilon_a$ is the photon acceptance of the prototype ARICH including the geometrical acceptance and detection efficiency of the HAPD, estimated to be around 42\% in the test, 
$\epsilon_q$ is the quantum efficiency of each HAPD, $\lambda$ is the wavelength of the Cherenkov photon and $\alpha$ is the fine structure constant.
In the beam test, the number of photoelectrons is estimated to be $\Npe = N_1 + N_2 = 2.787+ 7.969 = 10.756$.

\begin{figure}[ht]
	\centering
		\includegraphics[width=7cm]{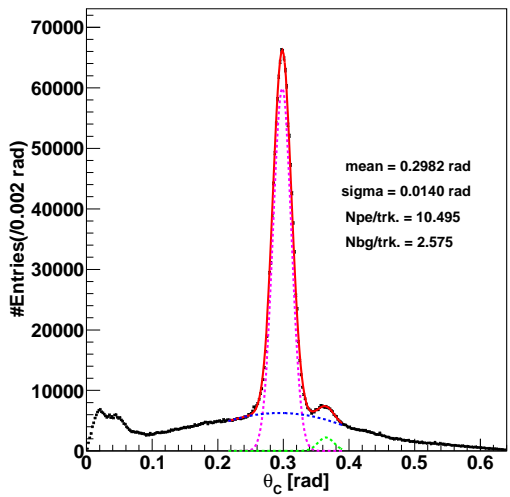}
	\caption{Cherenkov angular distribution.
	The solid line (red) represents the fitting function.
	The dotted lines represent the primary peak (magenta), the secondary peak (green) and background structures (blue).}
	\label{Cherenkov_dist}
\end{figure}

\Fig{Cherenkov_dist} shows the angular distribution of the Cherenkov light, which was preselected after the rejection of multiple track events.
The preselection was performed by fitting the distributions of number of photoelectrons using \Eq{Npe_fitter} for every \unit{0.002}{rad} of the Cherenkov angle distribution.
Here $A_1$ in \Eq{Npe_fitter} corresponds to the number of single-track events for each Cherenkov angle.

The Cherenkov emission angle $\thetaC$ is simply calculated by $\tan\thetaC = r/L$, where $r$ is the measured radius and $L$ is the distance between the quartz window of the HAPD and the averaged emission point in the aerogel.
Here, the averaged emission point is assumed to be the middle of the thickness of the upstream aerogel.

We obtained the average number of detected photoelectrons $\Npe$ from the area above the background form.
It amounts to $\Npe = 10.495 \pm 0.111$ per track.
The result is almost consistent with the expected $\Npe$ (= 10.756).

\subsection{Resolution of the Cherenkov angle}
The Cherenkov angle is calculated by fitting the angular distribution with a Gaussian for the primary Cherenkov peak, a second-order polynomial function for the background, and Gaussian for the small secondary peak seen at around \unit{\thetaC=0.36}{rad}, which is made by reflected photons on the APD surface.
Then we obtained the Cherenkov emission angle $\thetaC$ to be \unit{0.2982 \pm 0.0002}{rad}, and the angular resolution $\sigma_\theta$ to be \unit{14.03}{mrad} from the Gaussian assumption.
The obtained $\thetaC$ (\unit{0.2982 \pm 0.0002}{rad}) is compared with the expected $\thetaC$ calculated using \Eq{Cherenkov_fomula} as \unit{0.2998}{rad} assuming a \GeV{5} electron on $n=1.0467$ radiator.
We confirmed that the measured angle is as expected.

The main source of $\sigma_\theta$ is assumed to be $\sigma_\theta = \sqrt{\sigma_{\rm emp}^2 + \sigma_{\rm pix}^2}$ \cite{Matsumoto2004}, where $\sigma_{\rm emp}$ is the uncertainty in the emission point and is estimated to be $d\sin{\thetaC}/(L\sqrt{12})$, where $d$ (= \unit{20}{mm}) is the thickness of the aerogel and $L$ (= \unit{190}{mm}) is the distance between the averaged emission point in the aerogel and the surface of the HAPD, and $\sigma_{\rm pix}$ is the position resolution from the pixel size of the HAPD and is estimated to be $a\cos{\thetaC}/(L\sqrt{12})$, where $a$ (= \unit{4.9}{mm}) is the pixel size. 
We calculated $\sigma_\theta$ to be $(\sqrt{9.0^2 + 7.0^2} =)$ \unit{11.4}{mrad}. 
The measured resolution of the Cherenkov angle is \unit{14.03}{mrad} and is slightly different compared with the above estimation.
The discrepancy can account for around \unit{8}{mrad}, and is considered to arise from uncertainty related to the aerogel--- e.g., the effect of a non-flat surface and non-uniformities in the refractive index.
The effect of chromatic dispersion also makes the resolution worse but is negligible compared with the other effects~\cite{Matsumoto2004}.

The performance of the $\pi/K$ separation $S$ can be roughly estimated using the following equation:
\begin{eqnarray*}
	S = \frac{\Delta\thetaC}{\sigma_\theta}\sqrt{\Npe},
\end{eqnarray*}
where $\Delta\thetaC$ is the difference in Cherenkov angles between pion and kaon. 
$\Delta\thetaC$ is calculated to be \unit{23.7}{mrad} at \GeV{4}, as $\thetaC$ is calculated using \Eq{prin_ARICH} as \unit{0.2979}{rad} and \unit{0.2742}{rad} for a pion and a kaon, respectively.
Using the fit result of \fig{Cherenkov_dist} for $\thetaC$ (=\unit{14.03}{mrad}) and $\Npe$ (= 10.495), $S$ corresponds to \unit{5.47}{\sigma} of the $\pi/K$ separation.

\section{Particle identification efficiency}
To study the performance of the $\pi/K$ separation of the ARICH counter in a realistic situation, we perform event-by-event analysis based on the likelihood method for the beam test data.
We define probability density functions (PDFs) for the distributions of the Cherenkov angle and number of detected photoelectrons.
We prepare the PDFs for the signal and background assumptions.
Here, the signal and background are assumed to be pion and kaon, respectively, to emulate $\pi/K$ identification of the ARICH counter.
We calculate likelihoods for an event of the beam test.
We estimate the performance of the $\pi/K$ separation using likelihood ratio.

\subsection{Definition of the likelihood function}
As the first step, we define the likelihood function.
The likelihood function $\calL$ for an event is given by the following equation:
\begin{eqnarray}
	\calL &=& \calL_{\Npe} \times \calL_\theta,
	\label{Ltot}
\end{eqnarray}
where $\calL_{\Npe}$ is the likelihood of the number of detected photoelectrons in each event and $\calL_\theta$ is the likelihood of the Cherenkov angles for photoelectrons of an event.

\subsection{PDF construction}

The PDF for the number of detected photoelectrons per event is assumed to be a single Poisson distribution $Po(\mu)$.
The mean $\mu$ is quoted from the expected value calculated using \Eq{Npe_estimation} with the particle mass, the momentum, and the refractive indices of the aerogel layers as parameters.
For \GeV{3.5}, the expected number of detected photoelectrons is 10.629 and 8.938 for pion and kaon, respectively.

\begin{figure}[ht]
	\centering
	\includegraphics[width=15cm]{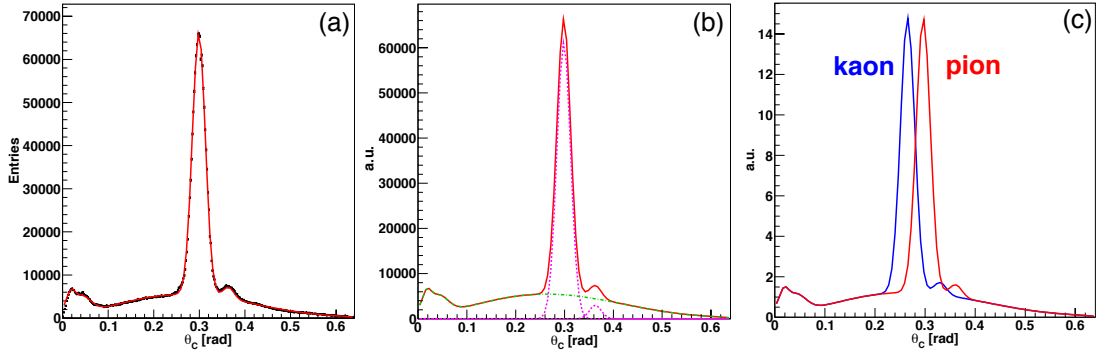}
	\caption{Construction scheme of the PDF for the Cherenkov angle.
	(a) Fitting into event selection applied data. The fitting function of four Gaussians and two eighth-order polynomials.
	(b) Parametrized distributions. The dotted lines (magenta) represent the Cherenkov signal peak. The dashed line (green) represents the common background. The solid line (red) represents the combined distribution.
	(c) PDF examples for pion and kaon at \GeV{3.5}.}
	\label{pdf_construction}
\end{figure}

Note that, we made the PDF of the Cherenkov angle for pion and kaon based on the electron data because the beam test is performed using the electron beam.
In order to generate the PDF of the Cherenkov angle as a function of momentum for the given mass assumption, we parametrized the Cherenkov angular distribution as a combination of some known functions.
The distributions for the primary Cherenkov peak and secondary peak made by the reflections of the photons at the APD surface are assumed to be Gaussian.
The slightly wide peak below \unit{0.1}{rad} originates from the Cherenkov photons, which were generated in the quartz window of the HAPD, and is assumed to be two Gaussians and an eighth-order polynomial.
The uncorrelated background is assumed to be an eighth-order polynomial.
Therefore, the entire distribution is fitted by four Gaussians and two eighth-order polynomial.
\Fig{pdf_construction}(a) shows the fitted distribution of the Cherenkov angle.

At the beam test, we obtained data by removing all the aerogels from the light-tight box in order to estimate the amount of background hits.
This data was fitted with a composite function, which had two Gaussians for the broad peak below \unit{0.1}{rad} and an eighth-order polynomial for the uncorrelated background.
\Fig{pdf_construction}(b) shows the components of the fitting function.
The dotted lines represent the primary Cherenkov peak and secondary peak.
A dashed line shows contributions from only the uncorrelated background.
A solid line is used to combine them, and it corresponds to the fitting function in \subfig{pdf_construction}{a}.
The primary Cherenkov peak and secondary peak for the background (kaon) assumption are shifted from the signal (pion) assumption by the Cherenkov angle difference between pions and kaons depending on the momentum.
The primary Cherenkov peak and secondary Cherenkov peak are added into the uncorrelated background to form the PDF.
The created PDFs for pion and kaon at \GeV{3.5} are shown in \subfig{pdf_construction}{c}.

The likelihood $\calL_\theta$ is given as the product of likelihoods for all the photoelectron hits in each event,
\begin{eqnarray}
	\calL_\theta &=& \prod_i^{\rm all\ hits} \calP_i(\theta),
	\label{Ltheta}
\end{eqnarray}
where $\calP_i(\theta)$ is a likelihood calculated using the PDF of the Cherenkov angle for $i$th hit in an event.

\subsection{Estimation of PID efficiency}
We estimate the PID performance of the ARICH counter for pion and kaon at \GeV{3.5} using single-track events taken with \GeV{5} electrons.
In order to select single-track events, we select data containing the number of detected photoelectrons below a cutoff value $N_{\rm cut}$ as the filled area in \fig{numhits}.
We set $N_{\rm cut}$ at 15.828 corresponding to $\mu_2 - 1.17\sigma_2$, where $\mu_2$ is the mean value of the Poisson distribution for double-track events ($2N + B$ = 21.217) using \Eq{Npe_fitter}, and $\sigma_2$ is $\sqrt{\mu_2}$.

\begin{figure}[ht]
	\centering
	\includegraphics[width=7cm]{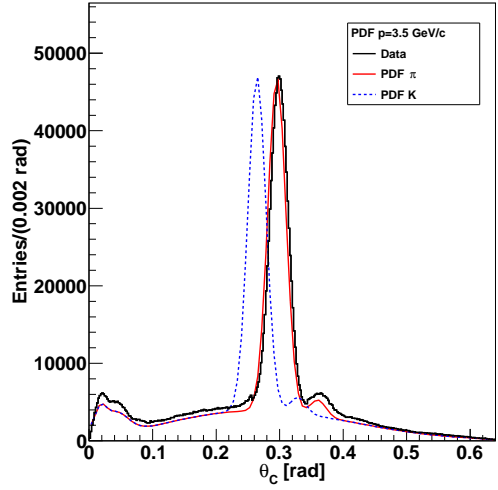}
	\caption{Cherenkov angular distribution of the beam test data with PDFs for pions and kaons at \GeV{3.5}.}
	\label{pdf_cangle}
\end{figure}

Because the expected Cherenkov angle $\thetaC(\pi)$ (= \unit{0.2973}{rad}) and expected number of photoelectrons $\Npe(\pi)$ (= 10.629) for \GeV{3.5} pions are close to expected $\thetaC(e)$ (= \unit{0.2998}{rad}) and expected $\Npe(e)$ (= 10.756) for \GeV{5.0} electrons, the selected data can be regarded as a data sample of \GeV{3.5} pions.
\Fig{pdf_cangle} shows the selected data distribution and PDFs of the Cherenkov angle for pion and kaon at \GeV{3.5}.

We define the likelihood ratio per event for pion $\calR(\pi)$ using the following equation in order to evaluate the PID performance.
\begin{eqnarray*}
	\calR(\pi) = \frac{\calL(\pi)}{\calL(\pi) + \calL(K)},
	\label{likelihood_ratio_pi}
\end{eqnarray*}
where $\calL(\pi)$ and $\calL(K)$ are the likelihoods for each particle.
These quantities were calculated for every event from \Eq{Ltot}.
We also define the likelihood ratio per event for kaons $\calR(K)$ using the following equation:
\begin{eqnarray*}
	\calR(K) = \frac{\calL(K)}{\calL(\pi) + \calL(K)} = 1 - \calR(\pi).
	\label{likelihood_ratio_K}
\end{eqnarray*}

\begin{figure}[ht]
	\center
	\subfigure[The difference of between the likelihood]{
		\includegraphics[width=7cm]{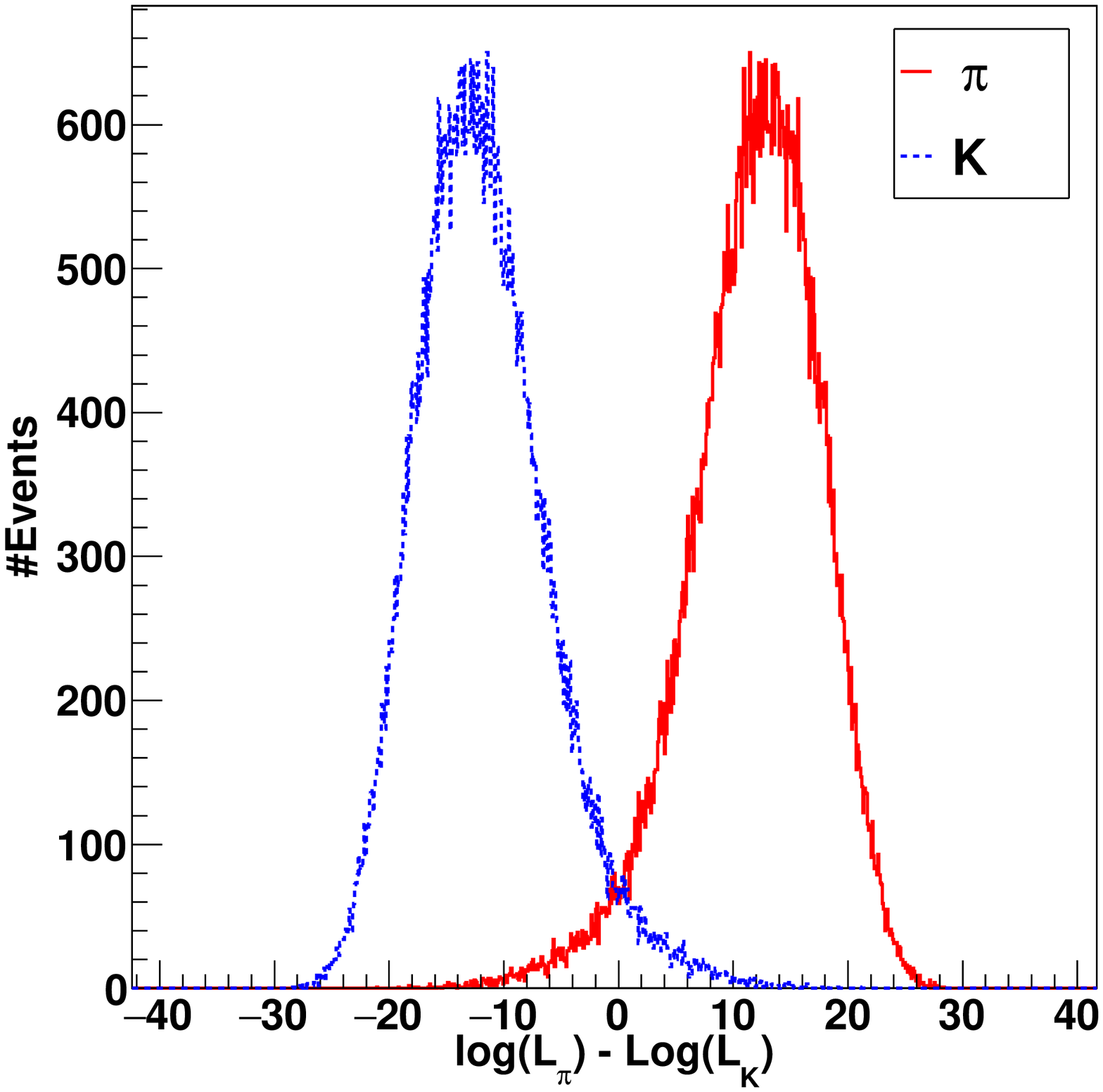}
	}
	\subfigure[Likelihood ratio distribution]{
		\includegraphics[width=7cm]{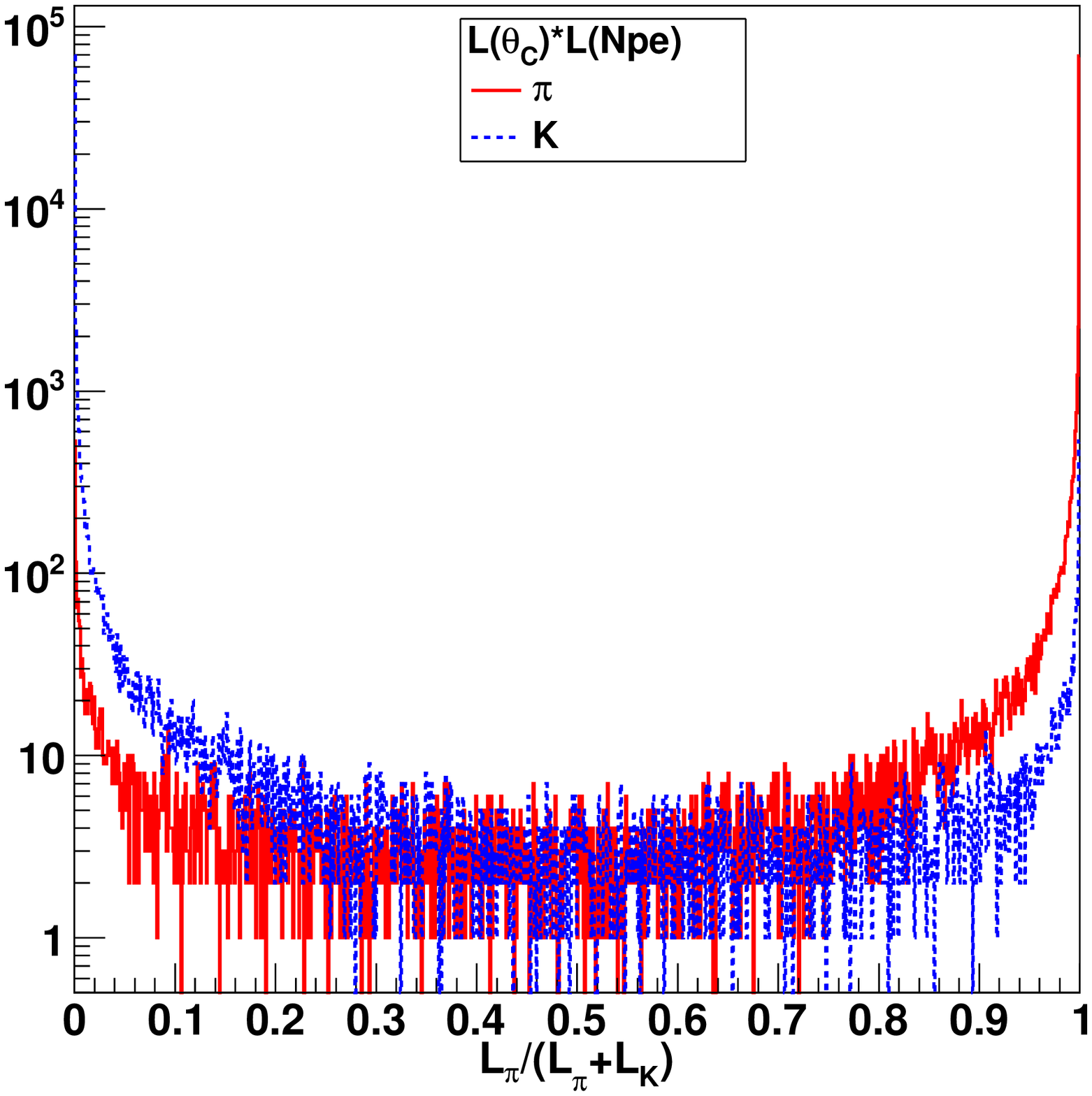}
	}	
	\caption{(a) Distribution of the likelihood difference between the pion (solid line) and kaon (dashed line) at \GeV{3.5}.
	(b) Likelihood ratio distribution for pions and kaons at \GeV{3.5}.}
	\label{likelihood_dist}
\end{figure}

\Fig{likelihood_dist}(a) shows the likelihood difference between the pion and kaon, which is calculated $\log{\calL(\pi)}-\log{\calL(K)}$.
\Fig{likelihood_dist}(b) shows the likelihood ratio $\calR(\pi)$ for the momentum assumption of \GeV{3.5}.
The solid and dotted lines represent the $\calR(\pi)$ and $\calR(K)$ respectively.

We define the $\pi$ identification efficiency for pion $\varepsilon(\pi)$ and kaon $\varepsilon(K)$ as the fraction of the number of events above the value $\calR_{\rm cut}$ and the number of total events.
It is equivalent to the following equation,
\begin{eqnarray*}
	\varepsilon(\pi) = \frac{{\rm \#Events} (\calR(\pi)>\calR_{\rm cut}) }{{\rm \#Events}({\rm All})}, \\
	\varepsilon(K) = \frac{{\rm \#Events} (\calR(K)>\calR_{\rm cut}) }{{\rm \#Events}({\rm All})}.
	\label{efficiency}
\end{eqnarray*}

When $\calR_{\rm cut}$ is set at 0.2, we obtained $\varepsilon(\pi)$ and $\varepsilon(K)$ as 97.4\% and 4.9\%, respectively.

%
%


%

\end{document}